\documentstyle[preprint,aps]{revtex}
\begin{document}
\draft
\title {New topologies in the phase diagram of the semi--infinite
Blume--Capel model}
\author {Carla Buzano and Alessandro Pelizzola}
\address {Dipartimento di Fisica and Unit\'a INFM,
Politecnico di Torino, I-10129 Torino, Italy}
\maketitle
\begin{abstract}
The phase diagram of the Blume--Capel model on a semi--infinite simple
cubic lattice with a (100) free surface is studied in the pair approximation
of the cluster variation method. Six main topologies are found, of which
two are new, due to the occurrence of a first order surface transition in
the phase with ordered bulk, separating two phases with large and small
surface order parameters. The latter is a new phase and is studied in some
detail, giving the behaviour of the order parameter profiles in two typical
cases. A comparison is made with the results of a low temperature
expansion, where these are available, showing a great increase in accuracy
with respect to the mean field approximation.
\end{abstract}
\vspace{1cm}
\pacs{PACS numbers: 05.50.+q \\[1cm] Keywords: cluster variation method,
Blume--Capel model, surface critical phenomena \\[1cm] Running title:
Semi--infinite Blume--Capel model}

\section{Introduction}

The Blume--Capel (BC) model \cite{blume,capel} is a spin--1 Ising model with
single--ion anisotropy that has been introduced as a model
for magnetic systems and then applied to multicomponent fluids
\cite{mukblu} and extended to the Blume--Emery--Griffiths model \cite{beg}
for He$^3$--He$^4$ mixtures. It is not exactly solvable in more than one
dimension, but it is has been studied over infinite $d$--dimensional
lattices by means of many different approximate techniques, and its phase
diagram is well known.

In recent years, when the theory of surface critical phenomena started
developing, some attention has been devoted to the study of the BC model
over semi--infinite lattices, with modified surface couplings. In particular
Benyoussef, Boccara and Saber \cite{bebosa} have determined the phase
diagram in the mean field approximation, reporting four possible topologies
at fixed bulk/surface coupling ratios, while Benyoussef, Boccara and el
Bouziani \cite{beboel} have done a similar analysis using a real space
renormalization group transformation. Other works, referring to
particular regions of the phase space, are those by Peliti and Leibler
\cite{pellei} and Crisanti and Peliti \cite{cripel} (real space
renormalization group), Jiang and Giri \cite{jiagir} (mean field
approximation), Tamura \cite{tamura} (effective field approximation) and
the present authors \cite{buzpel2,buzpel3} (CVM, low temperature expansion).

All these results show that, as in the case of the spin--1/2 Ising model
\cite{binder},
it is possible to have a phase with ordered surface and disordered bulk,
which is separated from the completely ordered
phase by the so--called {\em extraordinary} transition and
from the completely disordered phase by the {\em surface} transition.
When such a phase is absent, the transition between the completely ordered
and the completely disordered phase is named {\em ordinary} and the meeting
point of the lines of these three kinds of phase transitions is said {\em
special}, and is generally a multicritical point. While it is well
established that the ordinary transition can be either first or
second order, the possibility of a first order surface or extraordinary
transition has been
ruled out by Benyoussef, Boccara and el Bouziani \cite{beboel} and Tamura
\cite{tamura}, but not by the other authors.

In the present paper, which can be regarded as a refinement of a previous
work \cite{buzpel2}, we apply the pair approximation of the cluster
variation method (CVM) \cite{kik51,an,morita} to the Blume--Capel model
defined on a semi--infinite simple cubic lattice with a (100) free surface.
The dependence of the local quantities (order parameters, energy density,
and so on) on the distance from the free surface is taken into account by
dividing the lattice into an infinite set of layers parallel to the surface
and then treating the $N$th layer (we will use $N=5$) as if it was bulk.

Our results confirm the existence of first order extraordinary and surface
transitions (for the surface one we obtain very good agreement with a previous
low
temperature expansion analysis). Moreover, we show clearly that another first
order
transition can occur at the surface in presence of an ordered bulk, which
separates two phases with large and small values of the surface order
parameters, the latter being a new phase. When this
new transition is included in the phase diagram, the classification given
by Benyoussef, Boccara and Saber \cite{bebosa} is completed with the
addition of two new topologies.

The paper is organized as follows: in section II we define the model and
determine its ground state; in section III we
give the pair CVM free energy functional and describe the nested iteration
procedure we use to find its minimum; in section IV we describe our results,
giving the complete classification of the phase diagram topologies and
the behaviour of the order parameters and their profiles in the new phase.
Finally, in section V, we briefly summarize and discuss our results.

\section{The model and its ground state}

The BC model on a semi--infinite lattice $\Lambda$ with a free surface
denoted by $\partial \Lambda$ is defined by the (reduced) hamiltonian
\begin{equation}
\beta H = - J \sum_{\langle i j \rangle \not\subset \partial \Lambda} s_i s_j
- J_{\rm s} \sum_{\langle k l \rangle \subset \partial \Lambda} s_k s_l
+ \Delta \sum_{i \not\in \partial \Lambda} s_i^2
+ \Delta_{\rm s} \sum_{k \in \partial \Lambda} s_k^2 ,
\label{hbc}
\end{equation}
where $i,j,k,l$ are site labels, $s_i = \pm 1, 0$ is the $z$--component of a
spin--1
operator at site $i$, $J$ and $J_{\rm s}$ denote the bulk and surface
nearest neighbors (reduced) interactions (assumed positive),
respectively, $\Delta$ and
$\Delta_{\rm s}$ denote the bulk and surface single ion (reduced)
anisotropy and $\beta = (k_{\rm B} T)^{-1}$, with $k_{\rm B}$ Boltzmann's
constant and $T$ absolute temperature (from now on, we will express the
temperature in units of the bulk interaction, by setting $T \equiv 1/J$).
Furthermore, we introduce the coupling ratios $R = J/J_{\rm s}$ and $D =
\Delta/\Delta_{\rm s}$.

As it is often the case with classical discrete models, we can easily
determine the ground state of our model by looking for the lowest energy
configurations. This analysis is very important since it drives all
subsequent considerations.

For the bulk, as is well known, we have a state with broken
spin--flip symmetry for $\Delta < 3J$, given by e.g. $s_i = +1$ for all
bulk spins, and a state with unbroken symmetry, $s_i = 0$, for $\Delta >
3J$. These two states are associated with ordered and disordered bulk
phases, respectively, at finite temperature.

At the surface the situation is slightly more complex. Let us define the
states LO and HO by $s_i = 0, \forall i \in \partial \Lambda$ and
$s_i = +1, \forall i \in \partial \Lambda$, respectively. The corresponding
energies per site will be $E_{\rm LO} = 0$ and $E_{\rm HO} = \Delta_{\rm s}
- 2J_{\rm s} - m_0 J$, where $m_0 = 0$ in the bulk disordered phase and $1$
in the bulk ordered phase ($m_0$ is but the bulk order parameter $\langle
s_i \rangle$ at zero temperature).

We can now distinguish the following cases \cite{tesi}:
\begin{description}
\item{a)} when $D < 3R/(2+R)$ there is a surface transition at $\Delta/J =
\delta_0 \equiv D(2+R)/R$, in presence of an ordered bulk, and the states LO
and HO are
associated, at finite temperature, to two phases with ordered surface, but
with different values of the surface order parameters: the phase HO (we use
the same symbols for the phases and the corresponding pure states at $T=0$)
has large order parameters and is the usual ordered phase, while LO is a
new phase with small order parameters; this situation has
not been considered before, except in the limit $\Delta \to -\infty$, which
yields the spin--1/2 Ising model with a spin--1 free surface studied by
Kaneyoshi \cite{kaneyo}, in which a first order surface transition in the
presence of ordered bulk is obtained, but neither studied in detail nor
reported in the phase diagram;
\item{b)} when $3R/(2+R) < D < 3R/2$ there is a surface transition at
$\Delta = 3J$, in correspondence of the bulk transition, and the states LO
and HO are associated to phases with disordered and ordered surface,
respectively;
\item{c)} when $D > 3R/2$ there is a surface transition at $\Delta/J =
\delta_1 \equiv 2D/R$, in presence of a disordered bulk, and the states LO and
HO have the
same meaning as in the previous case.
\end{description}

\section{The CVM pair approximation for semi--infinite systems}

In order to study the finite temperature properties of the model, we
introduce now the pair approximation of the cluster variation method.

The CVM is based on an approximate expression of the entropy of the model
as a sum of contributions by all the elements of a set ${\cal M}$
consisting of certain
maximal clusters and all their subclusters. In the pair approximation, one
takes as maximal clusters all nearest neighbor pairs, and thus the
only subclusters to be considered are the sites. To each cluster $\gamma$
is associated a cluster (reduced) entropy given by
\begin{equation}
S_\gamma = - {\rm Tr}(\rho_\gamma \ln \rho_\gamma),
\label{cluentr}
\end{equation}
where $\rho_\gamma$ is a cluster density matrix, to be determined
minimizing the free energy. The total reduced entropy is then approximated by
\begin{equation}
S \simeq \sum_{\gamma \in {\cal M}} a_\gamma \nu_\gamma S_\gamma,
\label{totentr}
\end{equation}
where the coefficients $a_\gamma$ obey \cite{an,morita}
\begin{equation}
\sum_{\gamma \subseteq \alpha} a_\gamma = 1, \qquad \forall \alpha \in
{\cal M}
\label{agamma}
\end{equation}
and $\nu_\gamma$ represents the multiplicity (number of clusters per site)
of the cluster $\gamma$.

Since we have to deal with local quantities (density matrix elements, in
our formulation) which depend on the distance from the surface, we divide
our lattice, as mentioned in the introduction, into two dimensional layers
parallel to the surface, labeled by an integer $n$, $n=1$ being the surface
layer. Furthermore, to avoid working with an infinite number of variables,
we will treat the $N$th layer (typically we will choose $N = 5$) as if it
was bulk, by means of suitable constraints, and the bulk itself will be
considered in the CVM pair approximation, too, using the results obtained in
\cite{buzpel1}.

Following \cite{tesi}, we introduce the density matrices
$\rho_2^{(n)}(s_1,s_2)$ for the pairs having both sites in the $n$th layer,
$\rho_2^{(n,n+1)}(s_1,s_2)$ for those pairs having spin $s_1$ in
layer $n$ and spin $s_2$ in layer $n+1$, and $\rho_1^{(n)}(s_1)$ for sites
in layer $n$. The pair density matrices must satisfy the obvious constraints
\begin{equation}
\sum_{s_2} \rho_2^{(n)}(s_1,s_2) = \sum_{s_2} \rho_2^{(n-1,n)}(s_2,s_1) =
\sum_{s_2} \rho_2^{(n,n+1)}(s_1,s_2),
\label{compat}
\end{equation}
which ensure that the site density matrices can be properly defined, e.g. by
\begin{equation}
\rho_1^{(n)}(s_1) = \frac{1}{2} \sum_{s_2} \left[ \rho_2^{(n)}(s_1,s_2) +
\rho_2^{(n)}(s_2,s_1) \right].
\label{rho1n}
\end{equation}
Finally, in order to make the $N$th layer represent the bulk in an
effective way, we will treat the corresponding pair density matrix as a
constant, defined by
\begin{equation}
\rho_2^{(N)}(s_1,s_2) = \rho_{\rm bulk}(s_1,s_2),
\label{bulkconstr}
\end{equation}
where $\rho_{\rm bulk}$ is the normalized bulk pair density matrix given by
the CVM pair approximation \cite{buzpel1}. Notice that the normalization of
$\rho_{\rm bulk}$ and the constraints Eq.\ \ref{compat} imply the
normalization of all the density matrices we use.

Using the coefficients $a_\gamma$ and $\nu_\gamma$ reported in
\cite{buzpel2} one readily obtains the (reduced) free energy density
\begin{eqnarray}
f & = & - 2 J_{\rm s} {\rm Tr} (s_1 s_2 \rho_2^{(1)})
+ \Delta_{\rm s} {\rm Tr} (s_1^2 \rho_1^{(1)})
- J {\rm Tr} (s_1 s_2 \rho_2^{(1,2)}
\nonumber \\
&& + \sum_{n=2}^{N-1} \left[ - 2 J {\rm Tr} (s_1 s_2 \rho_2^{(n)})
+ \Delta {\rm Tr} (s_1^2 \rho_1^{(n)}) - J {\rm Tr} (s_1 s_2 \rho_2^{(n,n+1)})
\right] \nonumber \\
&& + \sum_{n=1}^{N-1} \left[ 2 {\rm Tr} (\rho_2^{(n)} \ln \rho_2^{(n)})
+ {\rm Tr} (\rho_2^{(n,n+1)} \ln \rho_2^{(n,n+1)})
+ a_1^{(n)} {\rm Tr} (\rho_1^{(n)} \ln \rho_1^{(n)}) \right] \nonumber \\
&& + \sum_{s_1} \sum_{n=1}^{N-1} \lambda_+^{(n)}(s_1)
\left[ \sum_{s_2} \rho_2^{(n,n+1)} (s_1,s_2) - \rho_1^{(n)}(s_1) \right]
\nonumber \\
&& + \sum_{s_1} \sum_{n=2}^{N} \lambda_-^{(n)}(s_1)
\left[ \sum_{s_2} \rho_2^{(n-1,n)} (s_2,s_1) - \rho_1^{(n)}(s_1) \right],
\label{free}
\end{eqnarray}
where $\lambda_\pm^{(n)}$ are Lagrange multipliers, $a_1^{(1)} = -4$
and $a_1^{(n)} = -5$ for $n > 1$.

Minimizing this functional corresponds to solving the following equations,
in the form of the natural iteration method \cite{kik74},
\begin{eqnarray}
\rho_2^{(1)}(s_1,s_2) & = & \exp \left[ \frac{1}{4} \left(
\lambda_+^{(n)}(s_1) + \lambda_+^{(n)}(s_2) \right) \right] \times
\nonumber \\
&& \times \exp\left[ J_{\rm s} s_1 s_2 - \frac{\Delta_{\rm s}}{4}(s_1^2 +
s_2^2)
\right] \rho_1^{(1)}(s_1)\rho_1^{(1)}(s_2)
\nonumber \\
\rho_2^{(n>1)}(s_1,s_2) & = & \exp \left[ \frac{1}{4} \left(
\lambda_+^{(n)}(s_1) + \lambda_+^{(n)}(s_2)
+ \lambda_-^{(n)}(s_1) + \lambda_-^{(n)}(s_2) \right) \right] \times
\nonumber \\
&& \times \exp\left[ J s_1 s_2 - \frac{\Delta}{4}(s_1^2 + s_2^2)
\right] \cdot \left[ \rho_1^{(n)}(s_1)\rho_1^{(n)}(s_2)
\right]^{5/4}, \nonumber \\
\rho_2^{(n,n+1)}(s_1,s_2) & = & \exp \left[ - \lambda_+^{(n)}(s_1)
- \lambda_-^{(n+1)}(s_2) \right] \exp \left( J s_1 s_2 \right),
\label{nim}
\end{eqnarray}
together with the minor iteration equations \cite{kik76,cvm3}
\begin{eqnarray}
\left[ \lambda_+^{(n)} (s_1) \right]_{r+1} = \left[ \lambda_+^{(n)} (s_1)
\right]_r + \frac{1}{2} \ln \frac{\displaystyle\sum_{s_2}
\rho_2^{(n,n+1)}(s_1,s_2)}
{\rho_1^{(n)}(s_1)} \nonumber \\
\left[ \lambda_-^{(n)} (s_1) \right]_{r+1} = \left[ \lambda_-^{(n)} (s_1)
\right]_r + \frac{1}{2} \ln \frac{\displaystyle\sum_{s_2}
\rho_2^{(n-1,n)}(s_2,s_1)}
{\rho_1^{(n)}(s_1)},
\label{mnr}
\end{eqnarray}
for the Lagrange multipliers, where $r$ and $r+1$ denote successive steps
of the iteration procedure.

The method of solution can be described as follows:
\begin{description}
\item[0.] initialize all pair density matrices
with a rough estimate of the solution;
\item[1.] determine the site density matrices according to Eq.\ \ref{rho1n};
\item[2.] recalculate the pair density matrices according to Eq.\ \ref{nim}
with all Lagrange multipliers set to 0;
\item[3.] solve Eq.\ \ref{mnr} for the Lagrange multipliers by simple
iteration;
\item[4.] determine the pair density matrices using Eq.\ \ref{nim} with the
values of the Lagrange multipliers obtained in step 3;
\item[5.] iterate steps 1--4 until the desired precision is reached.
\end{description}

Recalling that, in region of the phase space close enough to first order
transition lines, our free energy should have different minima
corresponding to stable and metastable phases, one has to repeat the above
procedure with different initializations, corresponding to the different
phase, and then choose the solution of minimum free energy. Once the
solution is found one readily obtains the order parameters
\begin{eqnarray}
m_n &=& \sum_{s = \pm 1, 0} s \rho_1^{(n)}(s) \\
q_n &=& \sum_{s = \pm 1, 0} s^2 \rho_1^{(n)}(s)
\label{ordpar}
\end{eqnarray}
and, in an analogous way, all the correlation functions.

\section{Phase diagram}

By means of the nested iteration procedure described in the previous
section, we have determined the phase diagram of our model in the plane
$(\Delta,T \equiv 1/J)$, for fixed values of $R$ and $D$. We have obtained
six main topologies, reported in Fig. 1, extending thereby the
classification by Benyoussef, Boccara and Saber \cite{bebosa}, in which cases
(a) and (d) were missing. These topologies are identified by specifying the
ground state, according to the classification given in section II, and the
result in the Ising limit $\Delta, \Delta_{\rm s}
\to -\infty$. Recalling the known results about the critical behaviour of
the semi--infinite Ising model \cite{binder} one realizes that in this
limit there are two possibilities: for $R > R_{\rm c}$ the model exhibits
an ordinary transition at the Ising bulk critical temperature, while for
$R < R_{\rm c}$ it undergoes an extraordinary transition at the bulk
critical temperature and a surface transition at a higher temperature. The
best estimate for $R_{\rm c}$ is about 0.66 \cite{lanbin,icm94}. In Fig.\ 1,
case (a)
corresponds to $R > R_{\rm c}$ and $D < 3R/(2+R)$, case (b) to $R > R_{\rm
c}$ and $3R/(2+R) < D < 3R/2$, case (c) to $R > R_{\rm c}$ and $D > 3R/2$,
case (d) to $R < R_{\rm c}$ and $D < 3R/(2+R)$, case (e) to $R < R_{\rm
c}$ and $3R/(2+R) < D < 3R/2$ and, finally, case (f) to $R < R_{\rm c}$
and $D > 3R/2$.

Notice that almost all of the above topologies, with the exception of cases
(b) and (f), could be further divided into subtopologies based on the order
of surface and bulk transitions at the special point $M$ and on the
eventual coalescence of that point with the critical point $C$. This is
exemplified in Fig. 2, which refers to cases (a) and (d) of Fig.\ 1.

It is also interesting to look at the behaviour of the order parameters in the
cases (a) and (d) of Fig.\ 1, where the new LO phase appears. In Figs. 3a and
3b,
which refer to case (a), we
see that the surface layer order parameters increase from zero at low
temperatures (it can be shown exactly that at low temperatures they behave
as $\exp(J - \Delta_{\rm s})$), while the order parameters of the inner
layers behave nearly as the bulk ones. Furthermore, in Figs. 4a and 4b,
the effect of the new surface phase transition is clearly shown.

Let us now turn our attention to the first order surface transition which
occurs in presence of a disordered bulk, which existence, questioned by
previous works \cite{beboel,tamura}, is clearly confirmed in our
approximation (Fig.\ 1, cases (c) and (f)). In Fig. 5 we have compared, for
one choice of $R$ and $D$, our results with those obtained by mean field
approximation \cite{bebosa} and low temperature expansion \cite{buzpel3}.
It is clearly seen that our approximation is a marked improvement with
respect to the simple mean field approximation, and the very good agreement
with the low temperature expansion results is a strong argument in favour
of the existence of such a transition.

Finally, we have not found any reentrant phenomenon in the phase diagram,
contrarily to previous results \cite{buzpel2}. This
discrepancy should be due to the very rough approximation used in
\cite{buzpel2}, where only one layer above the bulk was considered and, a
factor even more important, the method of solution adopted did not satisfy
the first of the constraints Eq.\ \ref{compat} in the case $n=N$ (=2, in
that case).

\section{Conclusions}

We have determined the phase diagram of the semi--infinite Blume--Capel
model using the pair approximation of the cluster variation method. Our
analysis has shown, at fixed $R$ and $D$, six main topologies, two of which
are completely new because of the appearance of a new ordered phase and,
consequently, of a new surface first order phase transition in presence of
an ordered bulk. This new phase has been studied in some detail and typical
behaviours of the order parameters profiles have been reported.
We have also given an example of how these main topologies may be slightly
modified by varying $R$ and $D$.

A question concerning the existence of the surface first order transition
in the presence of disordered bulk has been addressed and clarified by
comparing our results with previous results obtained by a low temperature
expansion and, finally, the issue of reentrant phenomena in the surface
transitions, raised in a previous paper, has been shortly discussed.

\vfill\eject

\begin{center}
{\bf FIGURE CAPTIONS}
\end{center}

\vspace{1cm}

\noindent FIG. 1. Phase diagram topologies: heavy and light lines denote bulk
and
surface transitions, respectively, while solid and dashed lines denote
second and first order transitions; $C$ and $M$ denote critical and
multicritical points.

\vspace{1cm}

\noindent FIG. 2. Phase diagram for $D = 0.2$ and $R = 1.00$, 0.50, 0.32
and 0.25.

\vspace{1cm}

\noindent FIG. 3a. Order parameters $m_n$ for $R = 1$, $D = 1/2$,
$\Delta/J = 1.6$; $n = 1$ (lowest curve), 2, 3, 4 and 5 (bulk, highest curve).

\vspace{1cm}

\noindent FIG. 3b. Order parameters $q_n$ in the case of Fig. 3a; $n = 1$
(lowest curve), 2, 3, 4 and 5 (bulk, highest curve).

\vspace{1cm}

\noindent FIG. 4a. The same as Fig. 3a for $R = 1/2$, $D = 1/2$, $\Delta/J =
2.6$.

\vspace{1cm}

\noindent FIG. 4b. The same as Fig. 3b for $R = 1/2$, $D = 1/2$, $\Delta/J =
2.6$.

\vspace{1cm}

\noindent FIG. 5. First order surface transition line for $R = 1$, $D = 1.53$
as given
by present method (solid line), mean field approximation (long dashes) and
low temperature expansion (short dashes).

\vfill\eject

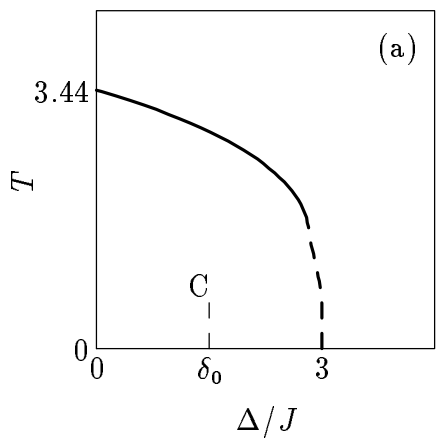
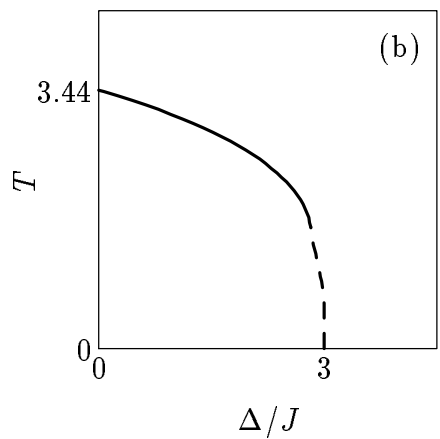
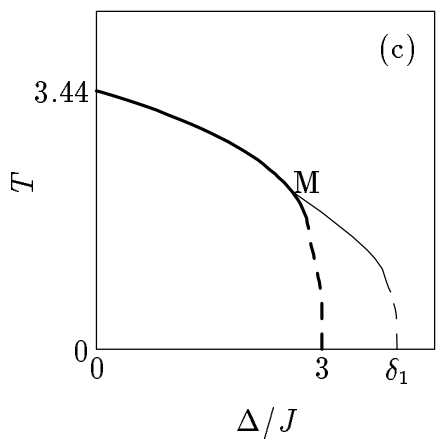
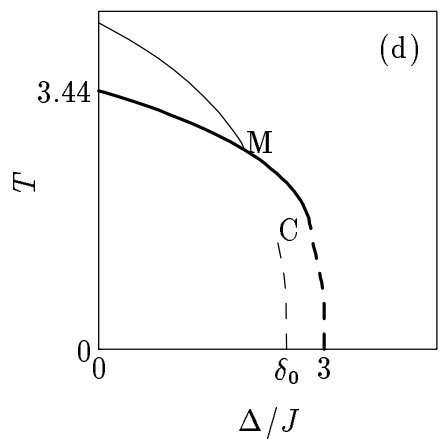
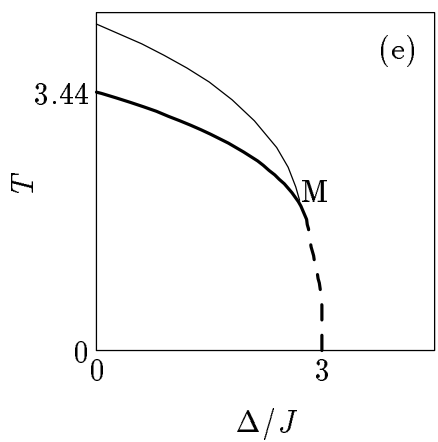
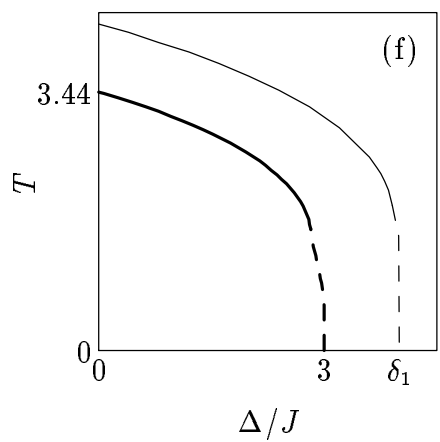
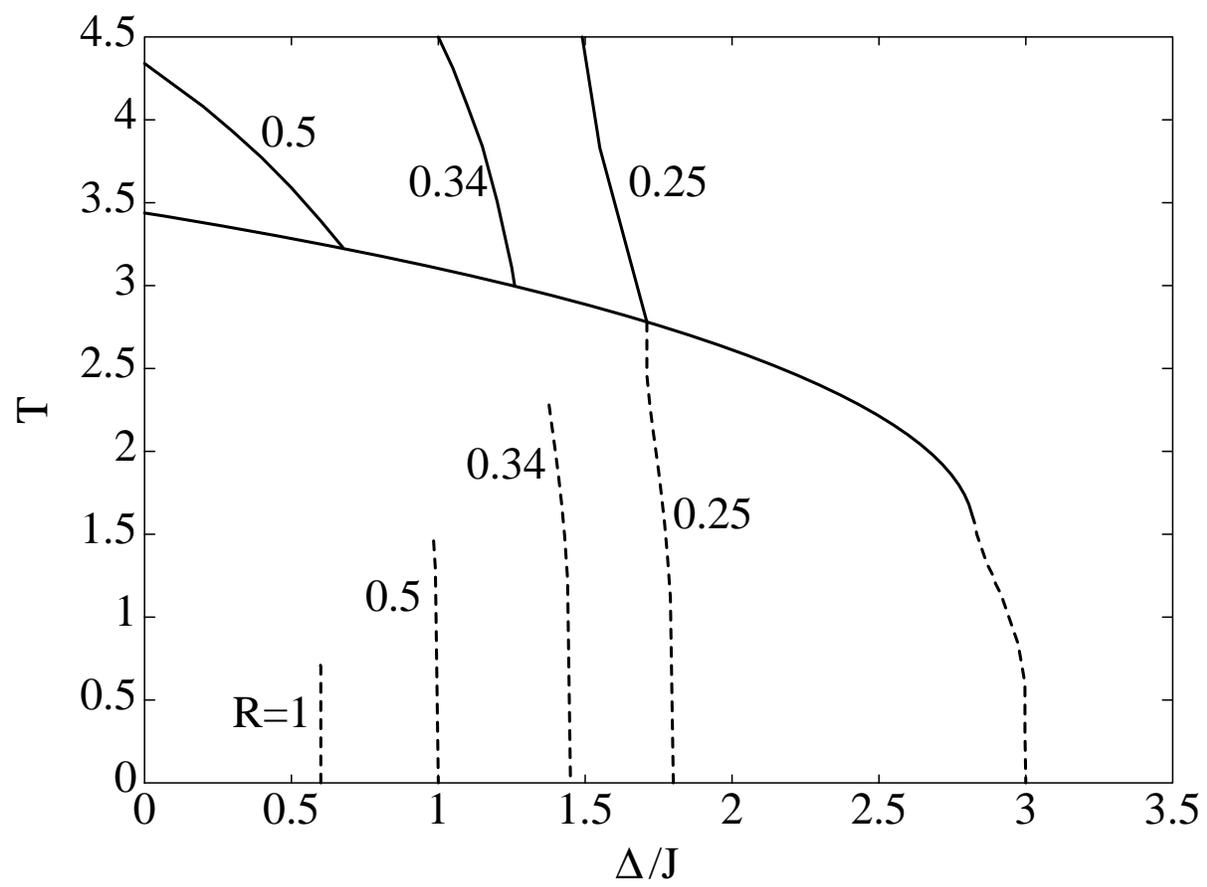
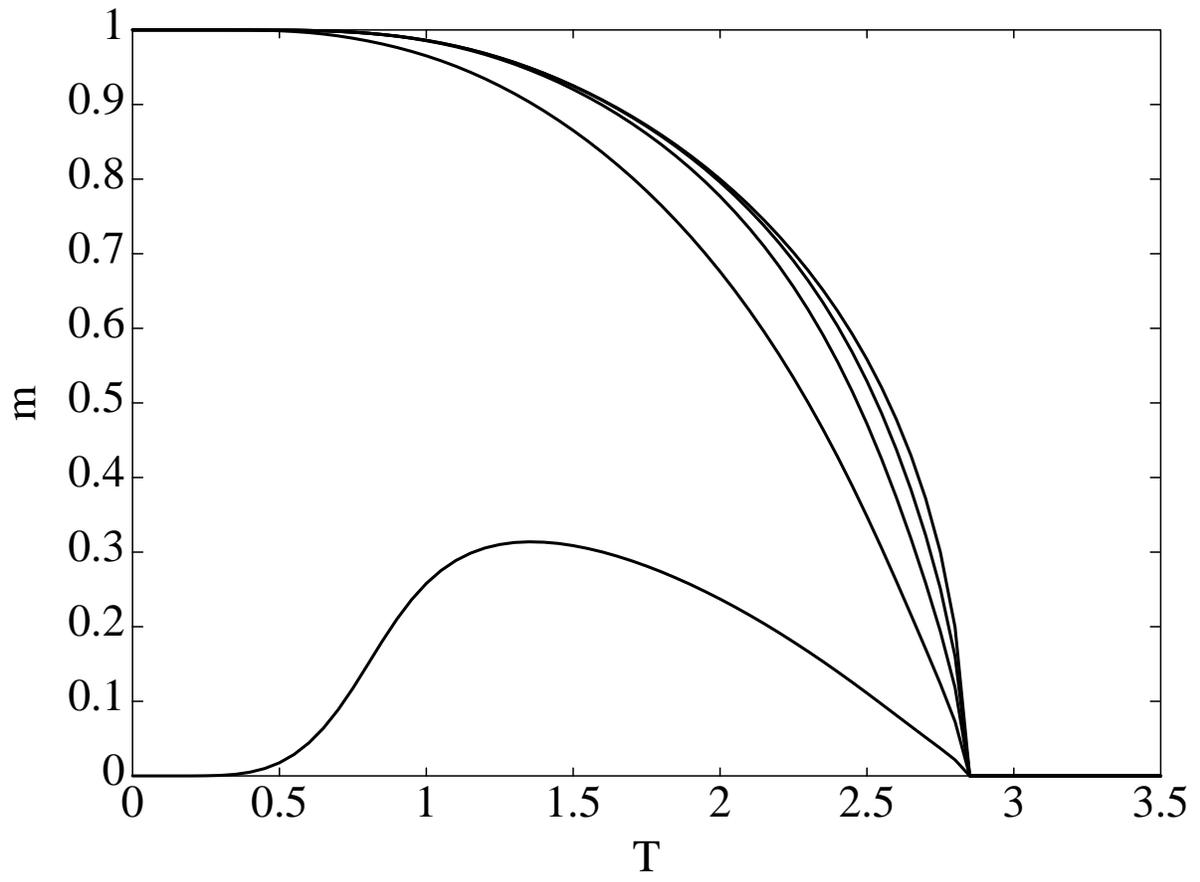
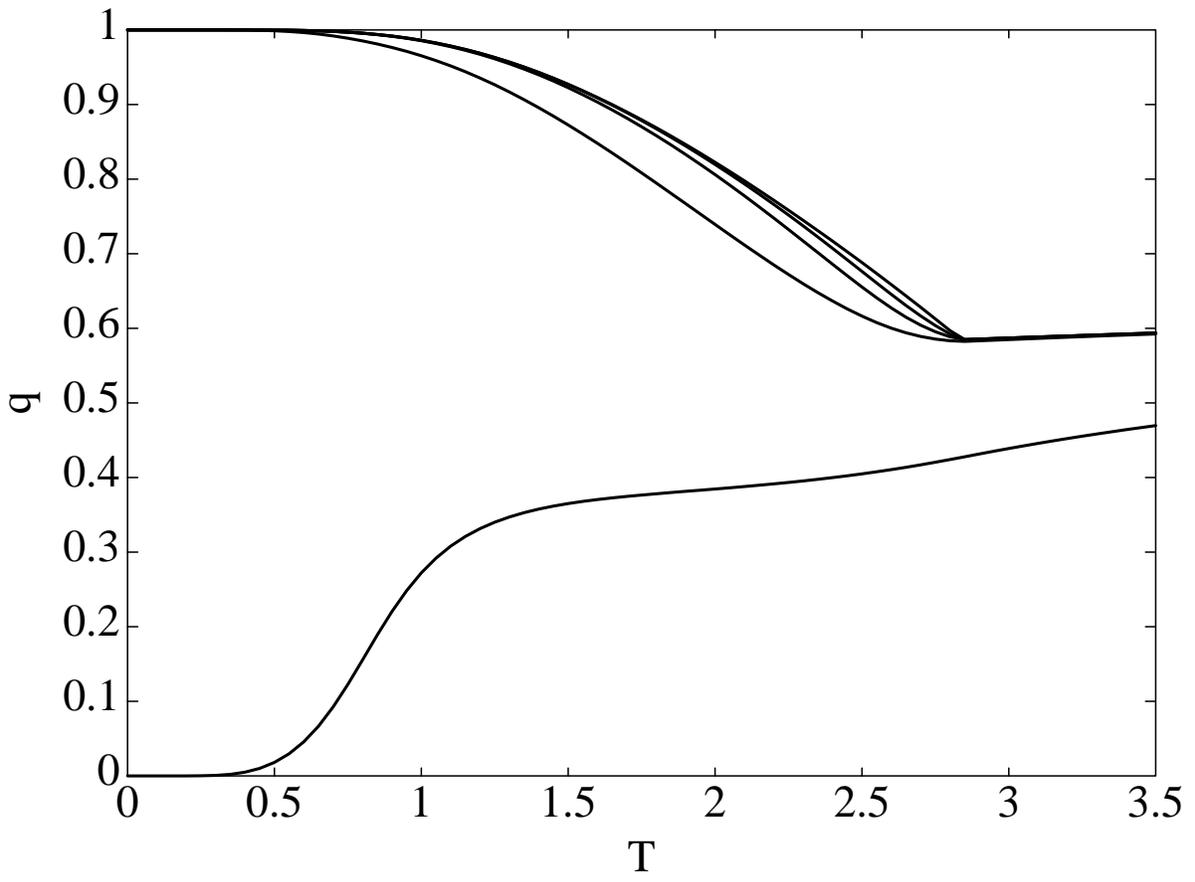
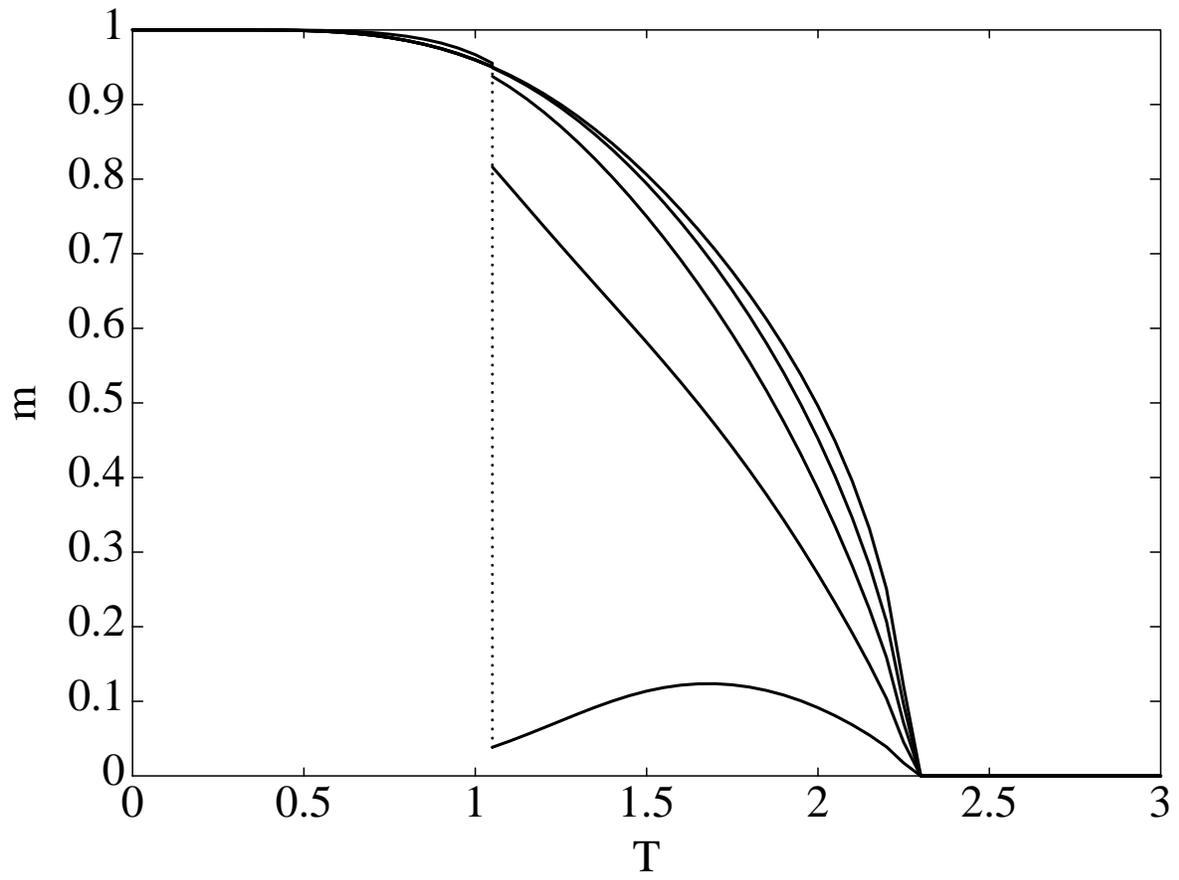
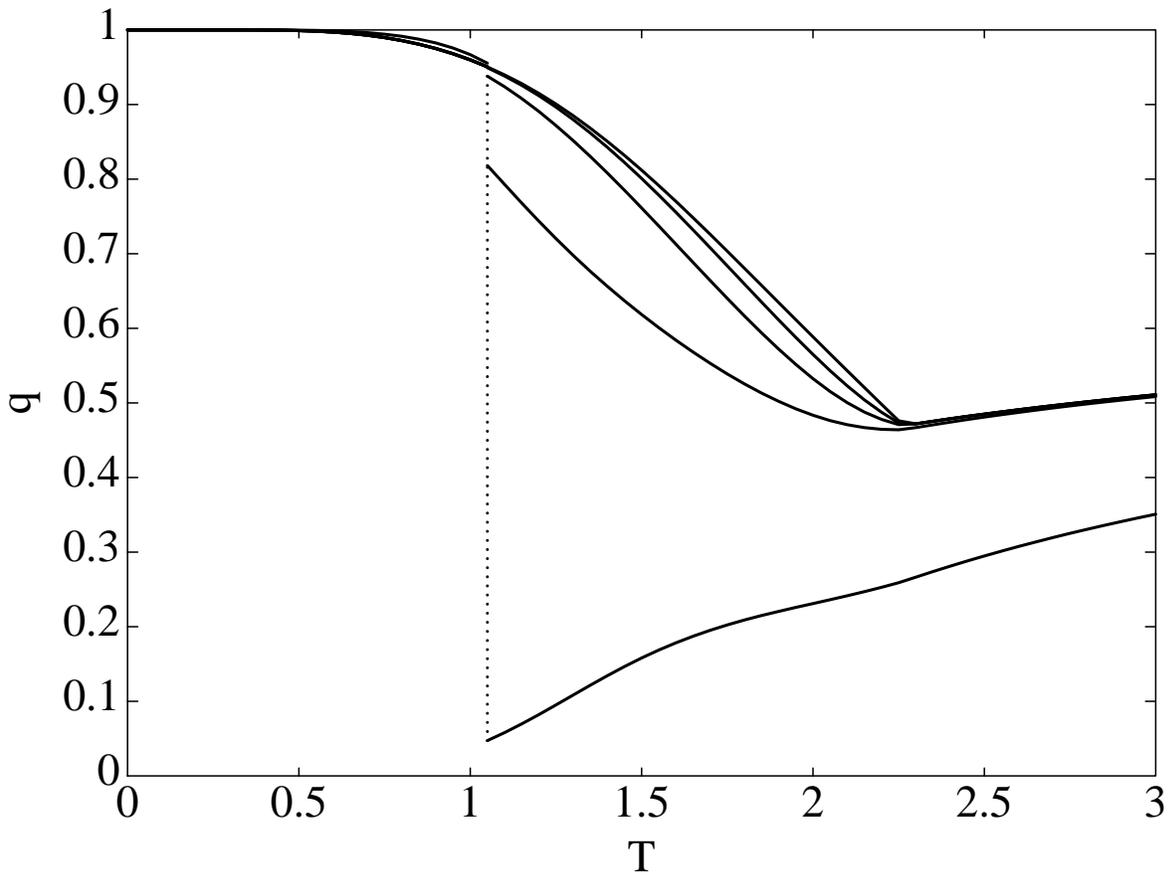
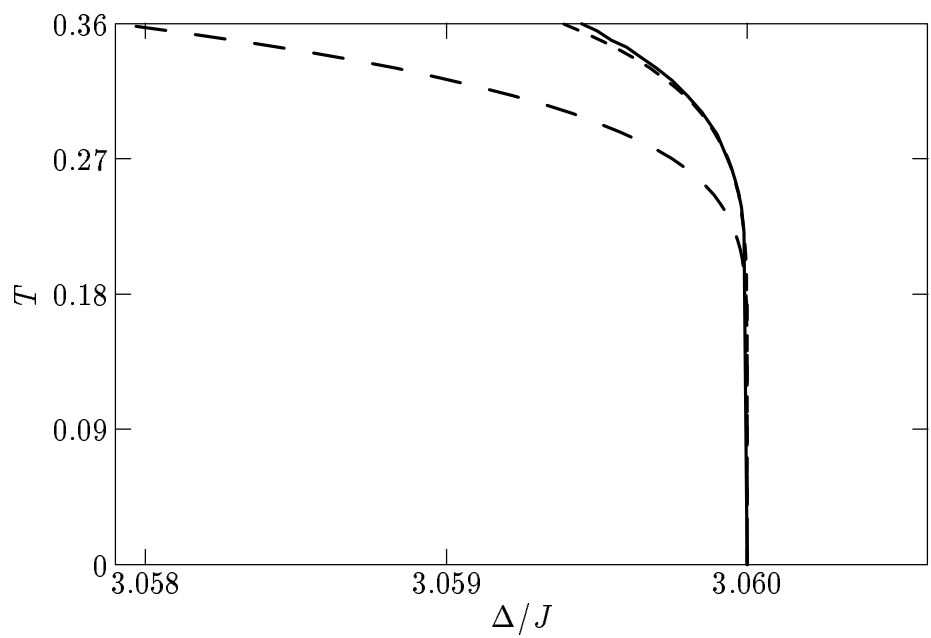
\end{document}